\documentclass[%
 superscriptaddress,
 reprint,% should use ``preprint'' here?
 preprintnumbers,
 amsmath, amssymb,
 aps,
 pra
]{revtex4-2}

\usepackage{%
	amsmath, 
	amssymb, 
	latexsym, 
	graphicx, 
	siunitx, 
	physics, 
	mathtools, 
	derivative, 
	mathrsfs,
    titlesec,
    titletoc,
    booktabs,
    nicematrix,
    array,
    threeparttable
}%
\usepackage[table]{xcolor}

\newcommand{\fig}[2]{Fig.~\ref{fig#1}(#2)}

\newcommand{\pdvv}[2]{\pdv[order={2}]{#1}{#2}}

\newcommand{\fsr}{\text{FSR}}
\newcommand{\figlabel}[1]{(#1)}
\newcommand{\figref}[1]{Fig.~\ref{#1}}
\newcommand{\Figref}[1]{Figure~\ref{#1}}
\usepackage{scalerel}
\DeclareRobustCommand{\Pinj}{P_{\scaleobj{0.9}{\textsl{inj}}}}

\DeclareRobustCommand{\Ainj}{A_{\scaleobj{0.9}{\textsl{inj}}}}
\DeclareRobustCommand{\Apump}{A_{\scaleobj{0.9}{\textsl{pump}}}}
\DeclareRobustCommand{\Ppump}{P_{\scaleobj{0.9}{\textsl{pump}}}}

\DeclareRobustCommand{\tR}{t_R}

\usepackage{dcolumn}
\usepackage[version=4]{mhchem}
\usepackage[hidelinks]{hyperref}% add hypertext capabilities
\usepackage{lipsum}
\usepackage[trueslanted]{newtxtext}
\begin{document}
	%\preprint{\textcolor{red}{DRAFT-\today}}
	%%%%%%%%%%%%%%%%%%%%%%%%%%%%%% TITLE %%%%%%%%%%%%%%%%%%%%%%%%%%%%%%%%%%%%%%%%%%%%%%%%%
	\title{All-optical Synchronization of Breather Solitons in a Kerr Microresonator}%
	
	\author{Wang Liao}
	\email{wang.liao@auckland.ac.nz}
	\affiliation{Department of Physics, The University of Auckland, Auckland 1142, New Zealand}
	\affiliation{The Dodd-Walls Centre for Photonics and Quantum technologies, Dunedin, New Zealand}
	
	\author{Stuart G. Murdoch}
	\affiliation{Department of Physics, The University of Auckland, Auckland 1142, New Zealand}
	\affiliation{The Dodd-Walls Centre for Photonics and Quantum technologies, Dunedin, New Zealand}
	
	\author{St\'ephane Coen}
	\affiliation{Department of Physics, The University of Auckland, Auckland 1142, New Zealand}
	\affiliation{The Dodd-Walls Centre for Photonics and Quantum technologies, Dunedin, New Zealand}
	
	\author{Gregory Moille}
	\affiliation{Joint Quantum Institute, NIST/University of Maryland, College Park, MD, USA}
	\affiliation{Microsystems and Nanotechnology Division, NIST, Gaithersbug, MD, USA}
	
	\author{Kartik Srinivasan}
	\affiliation{Joint Quantum Institute, NIST/University of Maryland, College Park, MD, USA}
	\affiliation{Microsystems and Nanotechnology Division, NIST, Gaithersbug, MD, USA}
	
	\author{Miro Erkintalo}
    \email{m.erkintalo@auckland.ac.nz}
	\affiliation{Department of Physics, The University of Auckland, Auckland 1142, New Zealand}
	\affiliation{The Dodd-Walls Centre for Photonics and Quantum technologies, Dunedin, New Zealand}
	 
	\date{\today}
	%%%%%%%%%%%%%%%%%%%%%%%%%%%%%%%%%%%%%%%%%%%%%%%%%%%%%%%%%%%%%%%%%%%%%%%%%%%%%%%%%%%%%%
	
	\begin{abstract}
		%%%%%%%%%%%%%%%%%%%%%% ABSTRACT %%%%%%%%%%%%%%%%%%%%%%%%%%%%%
		\begin{center}
			%\textcolor{red}{[\textsc{Abstract to be determined...}]}
			\begin{minipage}{0.9\textwidth}
			Microresonator Kerr solitons are promising candidates for the realization of miniaturized on-chip optical frequency combs.
			%They also display oscillatory instability to form breather solitons characterized by periodic variation of temporal profiles.
			For specific system parameters, these solitons are associated with oscillatory instabilities, leading to breathing dynamics characterized by periodically modulated temporal and spectral profiles. In this regime, the solitons form a frequency comb comprised of primary comb lines surrounded by sidebands separated by the breathing frequency.           
            Here, we numerically and experimentally demonstrate that the breathing sidebands can be all-optically synchronized to a weak monochromatic laser injected into the cavity, thus providing direct control of the soliton oscillation frequency. 
			%With the all-optical injection scheme, we characterize the locking behaviors in breather solitons and report the accompanied noise inhibition during synchronization. 
            We judiciously characterize the synchronization process, and show that it is accompanied by a strong reduction of noise in the soliton's breathing. Our results provide fundamental insights on oscillatory dissipative structures, and could enable new forms of composite optical frequency combs.
						\end{minipage}
		\end{center}
		%%%%%%%%%%%%%%%%%%%%%%%%%%%%%%%%%%%%%%%%%%%%%%%%%%%%%%%%%%%%%
	\end{abstract}
	\maketitle
	%%%%%%%%%%%%%%%%%%%%%%%%%%%%%%%%%%%%%% MAIN BODY TEXT %%%%%%%%%%%%%%%%%%%%%%%%%%%
	%\section*{Introduction}
	 \emph{Introduction}---Dissipative Kerr solitons (DKSs, also known as temporal cavity solitons~\cite{leo2010temporal}) generated in \mbox{high-Q} microresonators have emerged as an important technology over the past decade~\cite{herr2014temporal}. These solitons have paved the way to realize coherent and broadband frequency combs in chip-integrated platforms~\cite{weiner2017cavity} with proof-of-concept applications ranging from timekeeping~\cite{Newman:19,Wu2025VernierMicrocombs}, spectroscopy~\cite{Myoung2016, herr2019astrocombs}, ultra-fast optical ranging~\cite{trocha2018ultrafast, Xu:25}, to terabit capacity telecommunications~\cite{marin2017microresonator}.
		 
	 Besides their significant practical value, DKSs also display a rich variety of interesting nonlinear dynamics. They belong to the universal class of dissipative solitons and arise through the process of self-organization~\cite{grelu2012dissipative}. Moreover, for specific system parameters, DKSs are known to exhibit complex instability behaviors~\cite{leo2013dynamics, anderson_observations_2016}.
	 Oscillatory instability, whereby solitons maintain their localization but undergo periodic variations in their temporal profile~\cite{matsko2012excitation}, is a subject of particular interest. The resultant ``breather solitons'' have been observed in macroscopic fiber ring resonators~\cite{leo2013dynamics}, as well as monolithic microresonators~\cite{bao2016observation, lucas2017breathing, yu2017breather, bao2018observation}. They have also been observed in normal dispersion~\cite{bao2018observation} and multimode cavities~\cite{guo2017intermode}, and they are linked to other time-periodic structures in different physical systems~\cite{peng2019breathing, binder2000observation, remoissenet2013waves}. While the optical spectrum of a breather soliton still corresponds to a frequency comb, each ``major'' comb line is now surrounded by multiple ``minor'' frequency sidebands, or breathing sidebands, that are equally separated by the breathing frequency.

    Recently, there has been considerable interest in the ability to all-optically synchronize (phase lock) stable DKS frequency combs to an external reference~\cite{moille2023kerr, wildi2023sideband}. Here, a reference continuous wave (CW) laser is directly injected into the resonator to capture a single comb line of the soliton, resulting in entrainment and noise reduction of the comb repetition rate~\cite{Moille2025NoiseQuenching,shandilya2025all}. An interesting question that arises is: can the oscillation frequency of breathing solitons be similarly synchronized via external injection of CW laser light with frequency close to arbitrary breather sidebands? Previous studies have shown that the breathing frequency can be controlled via electro-optical pump modulation methods~\cite{wan2020frequency, papp2019kerr}, and by harnessing the intrinsic periodicity of the resonator~\cite{cole2019subharmonic}. However, to the best of our knowledge, an all-optical approach to attain passive synchronization and probing of synchronization around comb modes different from the pump has not been attempted so far. Addressing this question could enable the breather solitons' composite comb spectrum to find new applications (e.g., in metrology), and it would elucidate the fundamental nonlinear dynamics of oscillating dissipative structures.

	 In this work, we answer the question posed above and present a comprehensive experimental and numerical investigation of all-optical synchronization of breather solitons in Kerr microresonators. We show that a low-power laser injected sufficiently close to any breathing sideband in the frequency domain can capture that sideband, causing the breathing frequency to synchronize (or lock) to the injected laser. Our experiments are conducted on a chip-integrated silicon nitride microresonator with an approximately 1~THz free spectral range (FSR), and we demonstrate all-optical synchronization of breathing frequency across multiple comb lines and breathing sidebands. We characterize the synchronization behaviors, observing (i) direct evidence of noise reduction in breathing oscillations; (ii) previously unidentified locking asymmetries between high- and low-frequency breathing sidebands; and (iii) sympathetic shifts in comb line frequency when breathing sidebands are synchronized to the injection lasers.
	 We believe our work represents the first experimental demonstration and investigation of all-optical synchronization of breathing DKSs, providing a potential route to employing such solitons in technological applications and expanding our understanding of breather solitons in systems driven far from equilibrium.
	 
\begin{figure}[t]
	 	\centering
	 	\includegraphics{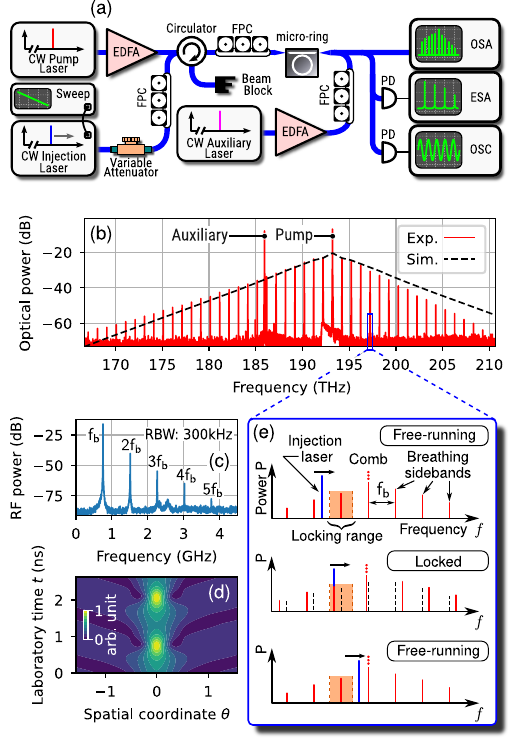}
	 	\caption{\label{fig1} \figlabel{a}~Experimental schematic. EDFA, erbium-doped fiber amplifier; FPC, fiber polarization controller; PD, photodiode; OSA, optical spectrum analyzer; ESA, electrical spectrum analyzer; OSC, oscilloscope. \figlabel{b}~Experimentally observed optical spectrum of a breathing soliton state (\textsl{red line}), and a corresponding simulated spectral profile (\textsl{black dashed line}).  \figlabel{c}~Measured RF spectrum of a breather soliton, showing harmonics separated by the breathing frequency $f_b$. \figlabel{d}~Simulated temporal evolution of a breather soliton over two breathing periods. \figlabel{e}~Zoom in section around a single comb line demonstrating the concept of breather synchronization: when the CW injection laser frequency enters the locking range of a breathing sideband, that sideband is captured and the breathing frequency is synchronized.}
	 \end{figure}

	 \emph{Results}---We begin by briefly describing our experimental setup shown in Fig.~\ref{fig1}(a).
	 Our on-chip microresonator is a micro-ring made of a 890~nm-thick \ce{Si3N4} waveguide with an outer diameter of \SI{46}{\micro\meter} and ring width of 1100~nm. It has an FSR of \SI{1006}{\giga\hertz} and a finesse of 3750.
	 The microresonator is driven by a pump derived from a CW external cavity diode laser (ECDL) at \SI{193.4}{\tera\hertz} (\SI{1550}{\nano\meter}).
	 After passing through an erbium-doped fiber amplifier (EDFA), the amplified \SI{500}{\milli\watt} pump enters the on-chip bus waveguide via a lensed-fiber. The light exiting the chip is then analyzed using a combination of optical and electrical spectrum analyzers. The thermal stability of the microresonator is maintained by using a \SI{186.2}{\tera\hertz} (\SI{1610}{\nano\meter}) counterpropagating CW auxiliary laser~\cite{Carmon:04, Zhang:19} that is polarized orthogonally with respect to the main pump in order to protect the pump mode family from the influence of the auxiliary laser \cite{zhou2019soliton}.
	  
	 We use the back-tuning technique described in~\cite{bao2016observation, lucas2017breathing} to access the breather soliton regime. Specifically, the pump laser wavelength is first set to the blue-detuned side of the resonance to observe the emergence of a ``primary'' modulation instability comb.
	 The wavelength is then increased to observe the transition into stable solitons. Once the stable soliton state is obtained, the pump laser wavelength is reduced until the characteristic oscillatory signature of a breather soliton appears. 

Figures~\ref{fig1}(b)-(d) demonstrate typical characteristics of breather solitons obtained from our experiments. Figure~\ref{fig1}(b) shows the soliton's optical frequency comb spectrum, spanning  from \SI{170}{\tera\hertz} to \SI{210}{\tera\hertz}, with a comb line separation of one FSR. The measured optical spectrum is in good agreement with results from numerical simulations (dashed black curve) that use the mean-field Lugiato-Lefever equation (LLE -- see Supplementary Information). The corresponding measured radio-frequency (RF) spectrum is shown in Fig.~\ref{fig1}(c), displaying strong harmonics at integer multiples of the breathing frequency, $f_b\approx 760~\mathrm{MHz}$, thus confirming oscillatory dynamics. (Of note: the measured breathing frequency agrees with the theoretical prediction that $f_\mathrm{b}\sim 1/(2\tau_\mathrm{c})$ with $\tau_\mathrm{c}$ the cavity photon lifetime.) Figure~\ref{fig1}(d) illustrates the evolution of the temporal profile of the breather soliton over two breathing periods obtained from numerical simulations.
	 
	 \begin{figure*}[t]
	 	\centering
	 	\includegraphics{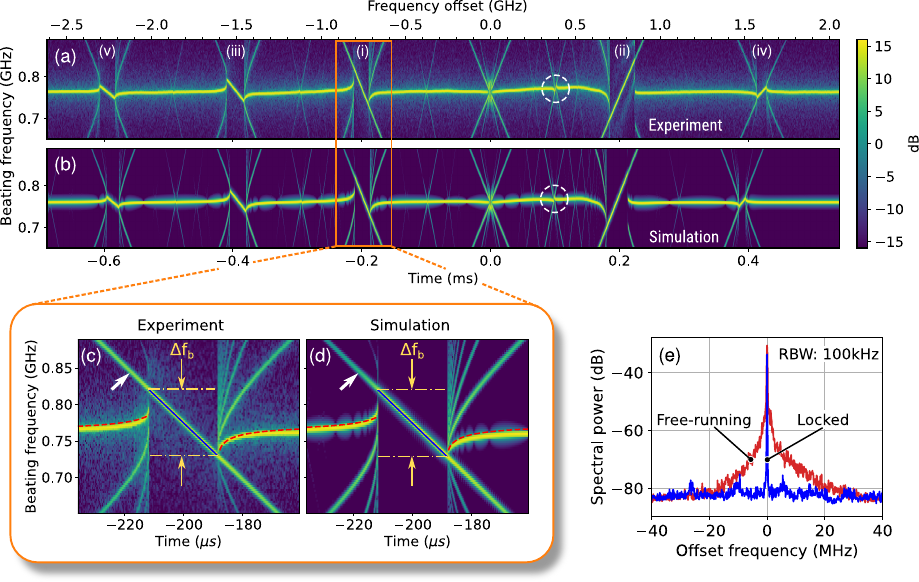}
	 	\caption{\label{fig2} Observation of breather synchronization near the pumped mode. \figlabel{a}~Measured RF spectrogram around the first breathing harmonic. When the injected laser sweeps across fundamental (highlighted by (i) and (ii)) or higher-order (highlighted by (iii)--(v)) breathing sidebands, synchronization happens. The frequency axis is estimated from the measured timebase considering a linear frequency ramp with a rate of \SI{3.8}{\giga\hertz/\milli\second}.    
        \figlabel{b}~Results from numerical simulations with parameters corresponding to experiments in (a).
	 	\figlabel{c},~\figlabel{d}~Zoom-in views of the measured and simulated synchronization of the fundamental low-frequency breathing harmonic, respectively. When the injected laser (marked by the white arrows) enters the locking range (between two orange dash-dotted lines), the first breather harmonic (red dashed line) exhibits synchronization (blue solid line). The red dashed curves follow the pulled breathing frequencies predicted from Adler theory (see main text) with free-running breathing frequency $f_\mathrm{b0} = 763~\mathrm{MHz}$ for time $t<-211~\mathrm{\mu s}$ and $f_\mathrm{b0} = 773~\mathrm{MHz}$ for time $t>-188~\mathrm{\mu s}$.
	 	\figlabel{e}~ RF spectrum around the first breathing harmonic measured on an ESA, showing that the linewidth shrinks significantly during synchronization (``Locked'') compared with the free-running spectral profile (``Free-running''), signalling noise reduction. RBW, resolution bandwidth.}
	 \end{figure*}
	 When in the breathing soliton state, each of the optical comb lines is surrounded by a sub-comb of ``breathing sidebands'' that are separated by the fundamental breathing frequency $f_b$. Our aim is to investigate whether, and to what extent, this breathing frequency can be all-optically controlled by injecting a secondary CW laser near one of the breathing sidebands (see \figref{fig1}(e)). To this end, attenuated CW light with power \mbox{$P_\mathrm{inj}\approx\SI{370}{\micro\watt}$} from a second independent ECDL (``CW Injection Laser'' in~\figref{fig1}(a)) is first configured to be co-polarized with respect to the primary pump laser and is then coupled into the microresonator. Next, we sweep the frequency of this secondary injection laser across the breathing sidebands while recording the resonator output power on a fast-photodiode (\SI{3}{\decibel} bandwidth of \SI{10}{\giga\hertz}). Finally, by applying a short-time Fourier transform (STFT) to the recorded power signals, we are able to check for signatures of breather synchronization. 
	 
	 \Figref{fig2}(a) shows a typical experimentally measured spectrogram, obtained from STFT analysis of the recorded power signal, when the frequency of the injected laser is ramped across the breathing sidebands around the pump mode (synchronization around non-pumped comb lines will be discussed later). Corresponding results obtained from numerical LLE simulations are shown in \figref{fig2}(b), and we see excellent agreement with the experimental result. The time and frequency axes of both spectrograms are zeroed at the point where the injected laser frequency coincides with that of the primary pump laser. In the immediate vicinity of this point, we observe a free-running breather oscillating at $(763.8 \pm 1.3)$~MHz. However, when the frequency of the injected laser is detuned from the pump by about \SI{\pm 760}{\mega\hertz} (corresponding to \SI{\pm 0.2}{\milli\second} in \figref{fig2}(a)), we observe clear signatures of synchronization: the breathing frequency abruptly changes and becomes entrained to the linear ramp of the injected laser frequency (highlighted by (i) and (ii) in \figref{fig2}(a)). This synchronization is maintained over a finite window of the injection laser frequency, beyond which the breathing frequency restores to the free-running value of about \SI{760}{\mega\hertz}. When the injection laser is further detuned, we observe an array of synchronization windows distributed regularly at intervals of free-running breathing frequencies (marked by (iii), (iv) and (v) in \figref{fig2}(a)). This indicates that the synchronization not only occurs for the fundamental breathing sidebands, but also for higher order sidebands.

    To provide further insights, Figs.~\ref{fig2}(c) and (d) show a zoomed-in view of the synchronization associated with the fundamental lower breathing sideband (i.e., the first sideband on the low frequency side of the pump mode). In the experimentally observed spectrogram~\fig{2}{c}, the beat signal between the injection laser and pump is indicated by the white arrow, and the first breathing harmonic signal is highlighted by the red dashed line.
	 When the injection laser enters the locking range indicated by the frequency span $\Delta f_\mathrm{b}$ (between two white dash-dotted lines in \fig{2}{c}),
	 we see that the beat signals of the injection laser and the breathing harmonic merge together to produce a single frequency trace (marked by the blue solid line). This shows that the injection laser has captured the low-frequency breathing sideband.
	 The spectrograms in Figs.~\ref{fig2}(c) and (d) also display typical features of injection locking in oscillators, including the transient frequency pulling between the locked and unlocked state~\cite{adler1946study}. In fact, the red dashed curves in Figs.~\ref{fig2}(c) and (d) appear to approximately follow the same functional form as oscillation frequencies derived from canonical Adler synchronization theory: $f_\mathrm{b} = f_\mathrm I \pm \sqrt{(f_\mathrm{I} - f_\mathrm{b0})^2 - (\Delta f_\mathrm{b}/2)^2}$, where $f_\mathrm b$ is the pulled breathing frequency, $f_\mathrm I$ is the injection laser frequency, and $f_\mathrm{b0}$ is the free-running breathing frequency.
     This observation can be theoretically understood by noting that the DKS breathing oscillations arise through a Hopf bifurcation~\cite{leo2013dynamics} that is canonically described by the Hopf normal form, from which the Adler synchronization equation can readily be derived by including an external drive term (see Supplementary Information). We note however that the agreement between Adler theory and our experiments and simulations is only qualitative: in obtaining the red-dashed curves in Fig.~\ref{fig2}(c) and (d), the free-running breathing frequency $f_\mathrm{b0}$ was modified to obtain a best fit (see figure caption). This highlights complexities beyond the simple Adler synchronization.

Close inspection of Figs.~\ref{fig2}(a) and (b) reveals that the breathing frequency is weakly synchronized sub-harmonically when the injection laser is detuned away from the pump by a fraction of the free-running breathing frequency. For example, the white circle in \figref{fig2}(a) highlights the sub-harmonic synchronization that occurs when then injection laser is detuned from the pump by half of the breathing frequency. We explain such sub-harmonic synchronization as a consequence of cascaded intracavity four-wave-mixing~\cite{erkintalo2012cascaded, xu2013cascaded}, where the nonlinear interaction between fields at the injected, pump, and breather sideband frequencies mix to generate idlers at $f_\mathrm{M,N} =  (f_\mathrm{p} + Mf_\mathrm{b}) + N(f_\mathrm{I} - f_\mathrm{P})$, where $M$ and $N$ are integers. If one these idlers is generated sufficiently close to a breathing sideband, synchronization can take place; it is straightforward to show that this occurs when $|f_\mathrm{p}-f_\mathrm{I}|$ is a rational fraction of the free-running breathing frequency, thus explaining the sub-harmonic synchronization seen in Fig.~\ref{fig2}. It is worth noting that the ability for a signal generated inside the resonator (via nonlinear wave mixing) to synchronize a breather is reminiscent of similar physics observed for the capture of primary comb lines of stable DKSs~\cite{Cole2023ChaoticGVHopping,Moille2025ParametricSync}.
	
    Our experiments also show signatures of noise reduction during the breather synchronization. This is evidenced in \figref{fig2}(e), which shows the RF spectral profiles of the first breathing harmonic measured on an electrical spectrum analyzer when the breather is free-running (red curve) and synchronized (blue curve).  Whilst a full characterization of the noise is beyond the scope of this Letter, the RF spectrum of the synchronized state is significantly narrower compared to the free-running case, signaling a quenching of the noise.
    
\begin{figure}
	 	\centering
	 	\includegraphics{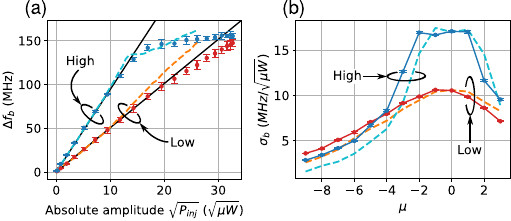}
	 	\caption{\label{fig3} 	 		%
	 		\figlabel{a} Synchronization range $\Delta f_b$ of fundamental breathing sidebands near the pump mode as a function of injection laser amplitude:
	 		red and blue dots are the experimental data for the lower and upper breathing sidebands, respectively, while
	 		orange and cyan dashed lines are corresponding simulation results.
            Black solid lines are linear fits to experimental data. 	 		\figlabel{b} Synchronization coefficient $\sigma_b$ (see main text for definition) as a function of the relative mode number, $\mu$. Solid curves are a guide to the eye, while other symbols have the same meaning as in (a). Error bars indicate $\pm1$ standard deviation computed from repeated measurements of the synchronization range, $\Delta f_b$}.
	 \end{figure}

	  We investigated how the  locking range (defined as the range of breather frequencies within the synchronization window, c.f. Fig.~\ref{fig2}(c)) depends upon the amplitude of the injected laser, $\sqrt{\Pinj}$. The red and blue dots (orange and cyan dashed lines) in Fig.~\ref{fig3}(a) show measured (LLE simulated) locking ranges when considering the synchronization of the fundamental low- and high-frequency breather sidebands around the pumped mode, respectively. Several conclusions can be drawn. First, in the low-value limit, the locking ranges increase linearly with the injection amplitude (in congruence with canonical Adler synchronization behavior): $\Delta f_\mathrm{b} = \sigma_\mathrm{b}\sqrt{\Pinj}$, where $\sigma_\mathrm{b}$ is a constant synchronization coefficient. Second, the linear trend breaks as the injection amplitude increases sufficiently. Third, the locking ranges for the low- and high-frequency sidebands show clear asymmetry: the synchronization coefficient $\sigma_\mathrm{b}$ for the latter is visibly larger than for the former. All of these features are observed both in our measured and simulated data, which show generally good agreement.

The asymmetric synchronization between the low- and high-frequency breather sidebands can be understood by recalling that DKSs (including breathers) are generated when the pump laser is red-detuned with respect to the cavity resonance. This naturally implies that the high-frequency breather sideband is closer to the (linear) cavity resonance than its low-frequency counterpart. Accordingly, for constant \emph{extracavity} injection laser amplitude $\sqrt{\Pinj}$, the amplitude of the \emph{intracavity} field at the injection laser frequency is larger when it is tuned close to the high-frequency breather sideband, resulting in more efficient synchronization.

All the results described above were obtained when the injection laser was tuned around the pumped mode. But of course, our all-optical synchronization scheme readily allows us to explore the salient dynamics around different comb lines (this is in contrast to electronic modulation schemes that would preclude such measurements due to the resonators' large FSR). To this end, we have repeated the measurements for all the comb lines accessible with two separate C-band and L-band injection lasers. We find that (i) synchronization occurs around all the comb lines when the injection laser is tuned close to a breather sideband, but (ii) the efficiency with which the synchronization occurs reduces as the absolute value of the relative mode number increase.  The latter point is highlighted in Fig.~\ref{fig3}(b), which shows measured synchronization coefficients $\sigma_b$ as a function of relative mode number, $\mu$, obtained by recording the locking ranges as a function of the injection power in the low-value limit (less than \SI{25}{\micro\watt}). Markers and solid curves show experimental results for both the fundamental low- and high-frequency sidebands. These results are in good agreement with corresponding results from LLE simulations (dashed curves).

The results in Fig.~\ref{fig3}(b) clearly demonstrate that the efficiency of breather synchronization is reduced around comb lines distant from the pump (in analogy with direct comb line injection locking~\cite{wildi2023sideband}.) Whilst this trend is observed for both the low- and high-frequency sidebands, it is interesting to note that their relative efficiency crosses over: the high-frequency sideband exhibits more efficient synchronization than its low-frequency counterpart for $\mu\approx 0$, but the roles reverse when $\mu<-4$. This observation can be understood by recalling that, in addition to the (linear) cavity resonance $\mathcal{C}$, the presence of a DKS endows the system with a secondary $\mathcal{S}$ resonance~\cite{guo2017universal}. Because of dispersion, the frequency offset between the $\mathcal{C}$ resonance and the comb line varies with the mode order $\mu$, causing the high-frequency breather sideband to be progressively further away in frequency from that resonance. In contrast, the $\mathcal{S}$ resonance (like the breather sidebands) exhibits a constant offset that overlaps with the low-frequency breather sideband. The ability to couple the injection laser into the resonator is thus comparatively less sensitive to the comb line order when it is injected close to low-frequency breather sideband rather than the high-frequency sideband.

We finally discuss the interplay between synchronization of breather sidebands and primary comb lines. The experiments described above measure the relative frequency difference between the primary comb (or pump) lines and the breather sidebands, rather than the absolute optical frequencies of any of the components. As such, the measurements do not distinguish shifts in the absolute optical comb lines (and thus the repetition rate). To overcome this shortcoming, we implement a heterodyne method to further probe the frequency components. Specifically, we mixed the light output from the cavity with a coherent fixed frequency laser as the local oscillator, and recorded all the beat tones between that oscillator and intracavity components.
    \begin{figure}
	 	\centering
	 	\includegraphics{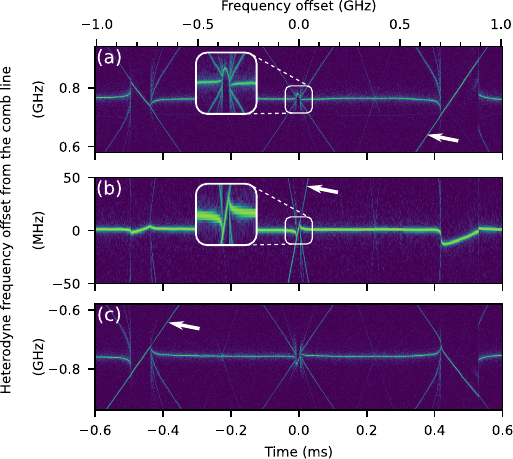}
	 	\caption{\label{fig4}%
        Heterodyne frequency beat tone spectrogram near the comb line at $\mu=-2$. The color scale follows the same color bar shown in~Fig.~\ref{fig2}. The time axis is zeroed at the point where the injection laser frequency coincides with the comb line. The white arrow in each panel marks the injection laser frequency component. The frequency offset axis is estimated from the measured timebase considering a linear frequency ramp with a rate of \SI{1.7}{\giga\hertz/\milli\second}.
        \figlabel{a} Frequency of the high-frequency breathing sideband. The inset shows the sympathetic shift in breathing frequency when the comb line is captured by the injection laser.
        \figlabel{b} Frequency of the $\mu=-2$ comb line. The sympathetic shifts in comb line frequency are obvious when breathing sidebands are captured by the injection laser. The inset shows the Kerr-induced comb line synchronization to the injection laser~\cite{moille2023kerr}. Note that the y-axis range is considerably smaller than in (a) and (c) for clarity.
        \figlabel{c} Frequency of the low-frequency breathing sideband.
        }
    \end{figure}
Figures~\ref{fig4}(a)--(c) show measured beat tone spectrograms when the injection laser is tuned around the $\mu=-2$ line. Breather synchronization described above is clearly visible from the measured data, but we also discern two new features. First, when the injection laser entrains a breather sideband, also the frequency of the comb line shifts, indicating a change in the comb repetition rate. Second, such sympathetic shifting is reciprocal: the breather frequency changes when the injection laser captures the comb line. These results are well reproduced by numerical simulations (see Supplementary Information), and they are suggestive of an intriguing interplay between the comb repetition rate (DKS group velocity) and breather oscillation that will warrant further investigation.

	 \emph{Discussion}---In conclusion, we have demonstrated that by injecting a secondary laser into a Kerr microresonator, the oscillatory behaviour of breathing DKSs can be controlled via optical synchronization. As is typical for injection locking, our experiments show clearly that the synchronization is associated with substantial reduction in the noise of breathing oscillation. Our all-optical injection scheme has enabled us to judiciously probe the synchronization phenomenon across multiple comb lines separated by 1~THz, revealing several non-trivial features, including sub-harmonic synchronization, asymmetries between low- and high-frequency breather sidebands, and dependence of synchronization on comb line order. Finally, by using optical heterodyne detection, we have probed for the interplay between synchronization of breather sidebands and primary comb lines, discovering that entrainment of one affects the other. All our measurements are in very good agreement with numerical simulations.  Our work opens up new avenues for controlling and stabilizing breather solitons, which could have practical applications (e.g. in metrology or optical computing~\cite{sun2025general, moy20221}) due to the states' composite comb structure. From a fundamental vantage, our work elucidates the interactions of breathing localized structures and plane waves in systems driven far from equilibrium.

	 \emph{Acknowledgments}---We acknowledge the financial support from the Marsden Funding of the Royal Society Te Ap\=arangi of New Zealand, with additional funding from the Dodd-Walls centre for Photonics and Quantum technologies. We thank Miles H. Anderson for insightful feedback. GM and KS acknowledge the NIST-on-a-chip program for its support.
     
	\bibliography{reference}

@article{zhou2019soliton,
  title={Soliton bursts and deterministic dissipative {Kerr} soliton generation in auxiliary-assisted microcavities},
  author={Zhou, Heng and Geng, Yong and Cui, Wenwen and Huang, Shu-Wei and Zhou, Qiang and Qiu, Kun and Wei Wong, Chee},
  journal={Light: Science \& Applications},
  volume={8},
  number={1},
  pages={50},
  year={2019},
  publisher={Nature Publishing Group UK London}
}

@article{leo2010temporal,
	title={Temporal cavity solitons in one-dimensional {Kerr} media as bits in an all-optical buffer},
	author={Leo, Fran{\c{c}}ois and Coen, St{\'e}phane and Kockaert, Pascal and Gorza, Simon-Pierre and Emplit, Philippe and Haelterman, Marc},
	journal={Nature Photonics},
	volume={4},
	number={7},
	pages={471--476},
	year={2010},
	publisher={Nature Publishing Group UK London}
}

@article{herr2019astrocombs,
  title={Astrocombs: recent advances},
  author={Herr, Tobias and McCracken, Richard A},
  journal={IEEE Photonics Technology Letters},
  volume={31},
  number={23},
  pages={1890--1893},
  year={2019},
  publisher={IEEE}
}

@article{Myoung2016,
	author = {Myoung-Gyun Suh  and Qi-Fan Yang  and Ki Youl Yang  and Xu Yi  and Kerry J. Vahala },
	title = {Microresonator soliton dual-comb spectroscopy},
	journal = {Science},
	volume = {354},
	number = {6312},
	pages = {600-603},
	year = {2016},
	doi = {10.1126/science.aah6516},
	URL = {https://www.science.org/doi/abs/10.1126/science.aah6516}
}

@article{herr2014temporal,
	title={Temporal solitons in optical microresonators},
	author={Herr, Tobias and Brasch, Victor and Jost, John D and Wang, Christine Y and Kondratiev, Nikita M and Gorodetsky, Michael L and Kippenberg, Tobias J},
	journal={Nature Photonics},
	volume={8},
	number={2},
	pages={145--152},
	year={2014},
	publisher={Nature Publishing Group UK London}
}

@article{marin2017microresonator,
	title={Microresonator-based solitons for massively parallel coherent optical communications},
	author={Marin-Palomo, Pablo and Kemal, Juned N and Karpov, Maxim and Kordts, Arne and Pfeifle, Joerg and Pfeiffer, Martin HP and Trocha, Philipp and Wolf, Stefan and Brasch, Victor and Anderson, Miles H and others},
	journal={Nature},
	volume={546},
	number={7657},
	pages={274--279},
	year={2017},
	publisher={Nature Publishing Group UK London}
}

@article{Newman:19,
	author = {Zachary L. Newman and Vincent Maurice and Tara Drake and Jordan R. Stone and Travis C. Briles and Daryl T. Spencer and Connor Fredrick and Qing Li and Daron Westly and B. R. Ilic and Boqiang Shen and Myoung-Gyun Suh and Ki Youl Yang and Cort Johnson and David M. S. Johnson and Leo Hollberg and Kerry J. Vahala and Kartik Srinivasan and Scott A. Diddams and John Kitching and Scott B. Papp and Matthew T. Hummon},
	journal = {Optica},
	number = {5},
	pages = {680--685},
	publisher = {Optica Publishing Group},
	title = {Architecture for the photonic integration of an optical atomic clock},
	volume = {6},
	month = {May},
	year = {2019},
	url = {https://opg.optica.org/optica/abstract.cfm?URI=optica-6-5-680},
	doi = {10.1364/OPTICA.6.000680}
}

@article{weiner2017cavity,
  title={Cavity solitons come of age},
  author={Weiner, Andrew M},
  journal={Nature Photonics},
  volume={11},
  number={9},
  pages={533--535},
  year={2017},
  publisher={Nature Publishing Group UK London}
}

@article{matsko2012excitation,
  title={On excitation of breather solitons in an optical microresonator},
  author={Matsko, AB and Savchenkov, AA and Maleki, L},
  journal={Optics letters},
  volume={37},
  number={23},
  pages={4856--4858},
  year={2012},
  publisher={Optica Publishing Group}
}

@article{moille2023kerr,
  title={{Kerr}-induced synchronization of a cavity soliton to an optical reference},
  author={Moille, Gr{\'e}gory and Stone, Jordan and Chojnacky, Michal and Shrestha, Rahul and Javid, Usman A and Menyuk, Curtis and Srinivasan, Kartik},
  journal={Nature},
  volume={624},
  number={7991},
  pages={267--274},
  year={2023},
  publisher={Nature Publishing Group UK London}
}

@article{wildi2023sideband,
  title={Sideband injection locking in microresonator frequency combs},
  author={Wildi, Thibault and Ulanov, Alexander and Englebert, Nicolas and Voumard, Thibault and Herr, Tobias},
  journal={APL photonics},
  volume={8},
  number={12},
  year={2023},
  publisher={AIP Publishing}
}

@article{adler1946study,
  title={A study of locking phenomena in oscillators},
  author={Adler, Robert},
  journal={Proceedings of the IRE},
  volume={34},
  number={6},
  pages={351--357},
  year={1946},
  publisher={IEEE}
}

@article{wan2020frequency,
  title={Frequency stabilization and tuning of breathing solitons in {{\ce{Si3N4}}} microresonators},
  author={Wan, Shuai and Niu, Rui and Wang, Zheng-Yu and Peng, Jin-Lan and Li, Ming and Li, Jin and Guo, Guang-Can and Zou, Chang-Ling and Dong, Chun-Hua},
  journal={Photonics Research},
  volume={8},
  number={8},
  pages={1342--1349},
  year={2020},
  publisher={Optica Publishing Group}
}

@article{Moille2025ParametricSync,
  author  = {Moille, Gr{\'e}gory and Shandilya, Pradyoth and Erkintalo, Miro and Menyuk, Curtis R. and Srinivasan, Kartik},
  title   = {On-Chip Parametric Synchronization of a Dissipative Kerr Soliton Microcomb},
  journal = {Physical Review Letters},
  year    = {2025},
  volume  = {134},
  number  = {19},
  pages   = {193802},
  doi     = {10.1103/PhysRevLett.134.193802},
  url     = {https://doi.org/10.1103/PhysRevLett.134.193802}
}

@article{Cole2023ChaoticGVHopping,
  author  = {Cole, Daniel C. and Lamb, Emily S. and Xue, Xianwen and Diddams, Scott A. and Papp, Scott B.},
  title   = {Toward chaotic group velocity hopping of an on-chip dissipative Kerr soliton},
  journal = {Optica},
  year    = {2023},
  volume  = {10},
  number  = {12},
  pages   = {1615--1622},
  doi     = {10.1364/OPTICA.496310},
  url     = {https://doi.org/10.1364/OPTICA.496310}
}

@article{Moille2025NoiseQuenching,
  author  = {Moille, Gr{\'e}gory and Shandilya, Pradyoth and Stone, Jordan and Menyuk, Curtis and Srinivasan, Kartik},
  title   = {All-optical noise quenching of an integrated frequency comb},
  journal = {Optica},
  year    = {2025},
  volume  = {12},
  number  = {7},
  pages   = {1020--1030},
  doi     = {10.1364/OPTICA.561954},
  url     = {https://doi.org/10.1364/OPTICA.561954}
}

@article{Wu2025VernierMicrocombs,
  author    = {Wu, Kaiyi and O'Malley, Nathan P. and Fatema, Saleha and Wang, Cong and Girardi, Marcello and Alshaykh, Mohammed S. and Ye, Zhichao and Leaird, Daniel E. and Qi, Minghao and Torres-Company, Victor and Weiner, Andrew M.},
  title     = {Vernier microcombs for integrated optical atomic clocks},
  journal   = {Nature Photonics},
  year      = {2025},
  volume    = {19},
  pages     = {400--406},
  doi       = {10.1038/s41566-025-01617-0},
  url       = {https://doi.org/10.1038/s41566-025-01617-0}
}

@article{leo2013dynamics,
  title={Dynamics of one-dimensional {Kerr} cavity solitons},
  author={Leo, Fran{\c{c}}ois and Gelens, Lendert and Emplit, Philippe and Haelterman, Marc and Coen, St{\'e}phane},
  journal={Optics express},
  volume={21},
  number={7},
  pages={9180--9191},
  year={2013},
  publisher={Optica Publishing Group}
}

@article{shandilya2025all,
	title={All-optical azimuthal trapping of dissipative Kerr multi-solitons for relative noise suppression},
	author={Shandilya, Pradyoth and Ou, Shao-Chien and Stone, Jordan and Menyuk, Curtis and Erkintalo, Miro and Srinivasan, Kartik and Moille, Gr{\'e}gory},
	journal={APL Photonics},
	volume={10},
	number={1},
	year={2025},
	publisher={AIP Publishing}
}

@article{bao2016observation,
  title={Observation of {Fermi-Pasta-Ulam} recurrence induced by breather solitons in an optical microresonator},
  author={Bao, Chengying and Jaramillo-Villegas, Jose A and Xuan, Yi and Leaird, Daniel E and Qi, Minghao and Weiner, Andrew M},
  journal={Physical review letters},
  volume={117},
  number={16},
  pages={163901},
  year={2016},
  publisher={APS}
}

@article{lucas2017breathing,
  title={Breathing dissipative solitons in optical microresonators},
  author={Lucas, Erwan and Karpov, Maxim and Guo, Hairun and Gorodetsky, ML and Kippenberg, Tobias J},
  journal={Nature communications},
  volume={8},
  number={1},
  pages={736},
  year={2017},
  publisher={Nature Publishing Group UK London}
}

@inproceedings{papp2019kerr,
  title={{Kerr}-breather-soliton classical time crystals},
  author={Papp, Scott B and Cole, Daniel C},
  booktitle={Nonlinear Optics},
  pages={NTu2A--4},
  year={2019},
  organization={Optica Publishing Group}
}

@article{cole2019subharmonic,
  title={Subharmonic entrainment of {Kerr} breather solitons},
  author={Cole, Daniel C and Papp, Scott B},
  journal={Physical review letters},
  volume={123},
  number={17},
  pages={173904},
  year={2019},
  publisher={APS}
}

@article{yu2017breather,
  title={Breather soliton dynamics in microresonators},
  author={Yu, Mengjie and Jang, Jae K and Okawachi, Yoshitomo and Griffith, Austin G and Luke, Kevin and Miller, Steven A and Ji, Xingchen and Lipson, Michal and Gaeta, Alexander L},
  journal={Nature communications},
  volume={8},
  number={1},
  pages={14569},
  year={2017},
  publisher={Nature Publishing Group UK London}
}

@article{peng2019breathing,
  title={Breathing dissipative solitons in mode-locked fiber lasers},
  author={Peng, Junsong and Boscolo, Sonia and Zhao, Zihan and Zeng, Heping},
  journal={Science advances},
  volume={5},
  number={11},
  pages={eaax1110},
  year={2019},
  publisher={American Association for the Advancement of Science}
}

@article{binder2000observation,
  title={Observation of breathers in~{Josephson} ladders},
  author={Binder, P and Abraimov, D and Ustinov, AV and Flach, S and Zolotaryuk, Ya},
  journal={Physical review letters},
  volume={84},
  number={4},
  pages={745},
  year={2000},
  publisher={APS}
}

@article{trocha2018ultrafast,
  title={Ultrafast optical ranging using microresonator soliton frequency combs},
  author={Trocha, Philipp and Karpov, Maxim and Ganin, Denis and Pfeiffer, Martin HP and Kordts, Arne and Wolf, S and Krockenberger, J and Marin-Palomo, Pablo and Weimann, Claudius and Randel, Sebastian and others},
  journal={Science},
  volume={359},
  number={6378},
  pages={887--891},
  year={2018},
  publisher={American Association for the Advancement of Science}
}

@book{remoissenet2013waves,
	title={Waves {Called} {Solitons:} {Concepts} and {Experiments}},
	author={Remoissenet, Michel},
	year={2013},
	publisher={Springer Science \& Business Media}
}

@article{Xu:25,
	author = {Yiqing Xu and St\'{e}phane Coen and Miro Erkintalo and Stuart G. Murdoch},
	journal = {Opt. Express},
	number = {3},
	pages = {4714--4724},
	publisher = {Optica Publishing Group},
	title = {Toward visible ultrafast imaging with a synchronously pumped switching wave {Kerr} frequency comb},
	volume = {33},
	month = {Feb},
	year = {2025},
	url = {https://opg.optica.org/oe/abstract.cfm?URI=oe-33-3-4714},
	doi = {10.1364/OE.551627}
}

@article{Zhang:19,
	author = {Shuangyou Zhang and Jonathan M. Silver and Leonardo Del Bino and Francois Copie and Michael T. M. Woodley and George N. Ghalanos and Andreas {\O}. Svela and Niall Moroney and Pascal Del'Haye},
	journal = {Optica},
	number = {2},
	pages = {206--212},
	publisher = {Optica Publishing Group},
	title = {Sub-milliwatt-level microresonator solitons with extended access range using an auxiliary laser},
	volume = {6},
	month = {Feb},
	year = {2019},
	url = {https://opg.optica.org/optica/abstract.cfm?URI=optica-6-2-206},
	doi = {10.1364/OPTICA.6.000206}
}

@article{Carmon:04,
author = {Tal Carmon and Lan Yang and Kerry J. Vahala},
journal = {Opt. Express},
number = {20},
pages = {4742--4750},
publisher = {Optica Publishing Group},
title = {Dynamical thermal behavior and thermal self-stability of microcavities},
volume = {12},
month = {Oct},
year = {2004},
url = {https://opg.optica.org/oe/abstract.cfm?URI=oe-12-20-4742}
}

@article{grelu2012dissipative,
  title={Dissipative solitons for mode-locked lasers},
  author={Grelu, Philippe and Akhmediev, Nail},
  journal={Nature photonics},
  volume={6},
  number={2},
  pages={84--92},
  year={2012},
  publisher={Nature Publishing Group UK London}
}

@article{bao2018observation,
  title={Observation of breathing dark pulses in normal dispersion optical microresonators},
  author={Bao, Chengying and Xuan, Yi and Wang, Cong and F{\"u}l{\"o}p, Attila and Leaird, Daniel E and Torres-Company, Victor and Qi, Minghao and Weiner, Andrew M},
  journal={Physical review letters},
  volume={121},
  number={25},
  pages={257401},
  year={2018},
  publisher={APS}
}

@article{guo2017intermode,
  title={Intermode breather solitons in optical microresonators},
  author={Guo, Hairun and Lucas, Erwan and Pfeiffer, Martin HP and Karpov, Maxim and Anderson, Miles and Liu, Junqiu and Geiselmann, Michael and Jost, John D and Kippenberg, Tobias J},
  journal={Physical Review X},
  volume={7},
  number={4},
  pages={041055},
  year={2017},
  publisher={APS}
}

@article{erkintalo2012cascaded,
  title={Cascaded phase matching and nonlinear symmetry breaking in fiber frequency combs},
  author={Erkintalo, Miro and Xu, YQ and Murdoch, SG and Dudley, John Micha{\"e}l and Genty, Go{\"e}ry},
  journal={Physical review letters},
  volume={109},
  number={22},
  pages={223904},
  year={2012},
  publisher={APS}
}

@article{xu2013cascaded,
  title={Cascaded {Bragg} scattering in fiber optics},
  author={Xu, YQ and Erkintalo, M and Genty, G and Murdoch, SG},
  journal={Optics Letters},
  volume={38},
  number={2},
  pages={142--144},
  year={2013},
  publisher={Optical Society of America}
}

@article{sun2025general,
  title={General oscillator-based {Ising-machine} models with phase-amplitude dynamics and polynomial interactions},
  author={Sun, Lianlong and Burns, Matthew X and Huang, Michael C},
  journal={Physical Review Applied},
  volume={24},
  number={4},
  pages={044035},
  year={2025},
  publisher={APS}
}

@article{moy20221,
  title={A 1,968-node coupled ring oscillator circuit for combinatorial optimization problem solving},
  author={Moy, William and Ahmed, Ibrahim and Chiu, Po-wei and Moy, John and Sapatnekar, Sachin S and Kim, Chris H},
  journal={Nature Electronics},
  volume={5},
  number={5},
  pages={310--317},
  year={2022},
  publisher={Nature Publishing Group UK London}
}

@article{guo2017universal,
  title={Universal dynamics and deterministic switching of dissipative {Kerr} solitons in optical microresonators},
  author={Guo, Hairun and Karpov, Maxim and Lucas, Erwan and Kordts, Arne and Pfeiffer, Martin HP and Brasch, Victor and Lihachev, Grigory and Lobanov, Valery E and Gorodetsky, Michael L and Kippenberg, Tobias J},
  journal={Nature Physics},
  volume={13},
  number={1},
  pages={94--102},
  year={2017},
  publisher={Nature Publishing Group UK London}
}

@book{kuznetsov2023elements,
  title={Elements of applied bifurcation theory},
  author={Kuznetsov, Yuri A},
  year={2023},
  publisher={Springer},
  series = {Applied Mathematical Sciences}
}

@book{dwight1961tables,
  title={Tables of integrals and other mathematical data},
  author={Dwight, Herbert Bristol},
  publisher={The Macmillan Company},
  year={1961}
}

@article{anderson_observations_2016,
    title = {Observations of spatiotemporal instabilities of temporal cavity solitons},
    volume = {3},
    issn = {2334-2536},
    url = {https://www.osapublishing.org/abstract.cfm?URI=optica-3-10-1071},
    doi = {10.1364/OPTICA.3.001071},
    number = {10},
    urldate = {2016-09-25},
    journal = {Optica},
    author = {Anderson, Miles and Leo, François and Coen, Stéphane and Erkintalo, Miro and Murdoch, Stuart G.},
    month = oct,
    year = {2016},
    pages = {1071},
}
\clearpage

\begin{widetext}

\section*{Supplementary Information}

  \section{Numerical simulation details}
  
    \noindent As described in the main manuscript, we are interested in all-optical synchronization of breather solitons in a dispersive Kerr resonator. All the numerical simulation results presented in that manuscript have been derived from the following Lugiato-Lefever equation (LLE):
   \begin{equation}\label{eq:supple-lle}
        t_R\pdv[]{A}{t} = \left( -\alpha -i\delta -iL\frac{\beta_2}{2}\pdvv{}{\tau} + iL\gamma |A|^2\right)A + \sqrt{\theta}A_\mathrm{pump} + \sqrt{\theta}A_\mathrm{inj}\exp{-i\Omega t-i\mu D_1\tau}.
    \end{equation}
    Here, $A = A(t, \tau)$ is the complex envelope of the intracavity electric field, with $t$ and $\tau$ the slow and fast time variables that describe the evolution of the field envelope at the scale of the cavity photon lifetime and over a single round trip, respectively. $\tR$ is the round trip time, $\alpha=\pi/\mathcal{F}$ represents half of the power lost per round trip with $\mathcal{F}$ the cavity finesse,  $\delta$ is the phase detuning, $L$ is the perimeter of the ring cavity, $\beta_2$ is the second-order dispersion coefficient, $\gamma$ is the Kerr nonlinear coefficient, $\theta$ is the coupling coefficient, $A_\mathrm{pump}$ is the pump field with units of $\sqrt{W}$, and $D_1 = 2\pi\fsr = 2\pi/\tR$, with $\text{FSR}$ the cavity free-spectral range. $A_\mathrm{inj}$ represents the complex amplitude of the secondary injection field that is coupled into the cavity to all-optically synchronize breather solitons, with $\mu$ the relative mode number that is closest in frequency to the injection field and $\Omega$ the angular frequency separation between the injection frequency, $\omega_\mathrm{inj}$ and the $\mu$-th free-running comb line, $\omega_\mu$, i.e., $\Omega = \omega_\mathrm{inj}-\omega_\mu$.

    We numerically simulated the time evolution of the intracavity field envelope, $A(t, \tau)$, by integrating \eqref{eq:supple-lle} using the split-step Fourier method over the fast-time domain $(-t_R/2, t_R/2]$.  We set the initial condition of our numerical simulations to correspond to the approximate analytical soliton solution of the LLE given by 
    \begin{equation}
        A(t=0, \tau) = \sqrt{\frac{2\delta}{L\gamma}}\sech{\left(\tau\sqrt{\frac{2\delta}{L|\beta_2|}}\right)},\text{where }-t_R/2<\tau\leqslant t_R/2.
    \end{equation}
    To model the synchronization behaviors when sweeping the injection laser frequency, we let the frequency separation $\Omega$ be time dependent: $\Omega = \Omega(t)$ in \eqref{eq:supple-lle}. For a linear frequency ramp of the injection laser at a constant sweeping speed $\mathfrak a$, we set the frequency separation to be $\Omega(t) = \omega_{\text{I, }t=0} + \mathfrak at/2 - \omega_\mu$ such that the instantaneous frequency calculated by taking the slow time derivative of $\Omega(t)t$ changes at the same constant sweeping speed~$\mathfrak a$.

    All numerical simulations share the same set of system parameters which are given in Table~\ref{tab:supple-parameters across simulations}.
    \begin{table}[h]
        \begin{center}
        \begin{NiceTabular}{cc}
        \toprule[1pt]
            System parameter & Value\\
             \midrule[0.5pt]
             $\tR$ & \SI{0.994}{\pico\second}\\
             \rowcolor{gray!20} $\alpha$ & \SI{8.369e-4}{} \\             
             $L$ & $\pi\cdot\SI{46}{\micro\meter}$\\
             \rowcolor{gray!20} $\beta_2$ & \SI{-69.390}{\pico\second^2/\kilo\meter}\\
             $\gamma$ & \SI{2.737}{\per\watt\per\meter}\\
             \rowcolor{gray!20} $\theta$ & \SI{4.243e-4}{}\\
             \bottomrule[1pt]
        \end{NiceTabular}
        \end{center}
        \caption{Shared system parameters across all simulations}
        \label{tab:supple-parameters across simulations}
    \end{table}
    
    The variable parameters that are pertinent to each individual simulation discussed in the main text are the phase detuning $\delta$, the pump power $\Ppump = |\Apump|^2$, and the injection laser power $\Pinj=|\Ainj|^2$. We summarize the values of these parameters in Table~\ref{tab:supple-parameters individual simulation} according to the figure labels.
    \begin{table}[hb]
            \begin{threeparttable}
                \begin{center}
                    \begin{NiceTabular}{cccc}
                    \toprule[1pt]
                         Figure label in the main text & $\delta$ & $\Ppump$ & $\Pinj$\\
                         \midrule[0.5pt]
                                            Figure 1(b)\tnote{1} and (d) & \SI{5.270}{\milli\radian}  & \SI{47.14}{\milli\watt} & \SI{0.00}{\micro\watt}          \\
                         \rowcolor{gray!20} Figure 2(b)          and (d) & \SI{5.352}{\milli\radian}  & \SI{50.18}{\milli\watt} & \SI{56.15}{\micro\watt}          \\
                                            Figure 3(a)\tnote{2}         & \SI{5.444}{\milli\radian}  & \SI{53.82}{\milli\watt} & $\leqslant \SI{212.11}{\micro\watt}$\\
                         \rowcolor{gray!20} Figure 3(b)\tnote{2}         & \SI{5.270}{\milli\radian}  & \SI{47.14}{\milli\watt} & $\leqslant \SI{11.55}{\micro\watt}$\\
                         \bottomrule[1pt]
                    \end{NiceTabular}
                    \begin{tablenotes}
                    \centering
                        \item[1] Exponential moving average was applied to obtain a static spectral profile.
                        \item[2] The loss at the coupling between the lensed fiber and the chip endface was taken into account.
                    \end{tablenotes}
                \end{center}
            \end{threeparttable}
        \caption{System parameters used in individual simulations}
        \label{tab:supple-parameters individual simulation}
    \end{table}
    In addition, we applied an exponential moving average of the dynamic spectrum with a smoothing factor \SI{1e-3}{} to obtain the static spectral profile shown in Fig.~1(b) of the main text. Furthermore, for the simulation shown in Fig.~3(a) of the main text, we first compared the numerical and experimental results for the lower frequency breathing sideband to obtain the optical loss at the coupling point between the lensed fiber and the chip endface, and then we corrected the higher frequency breathing sideband numerical result according to the computed fiber-endface coupling loss. This correction procedure has been applied to the numerical simulation result shown in Fig.~3(b), where the loss factor was computed by comparing the numerical and experimental results for the breathing sideband at the low frequency side of the pump mode.
    
    To numerically simulate the experimental observation shown in Fig.~4 in the main text, we set the phase detuning $\delta = \SI{5.390}{\milli\radian}$, the pump power $\Pinj = \SI{51.71}{\milli\watt}$, and the injection laser power $\Pinj = \SI{79.38}{\micro\watt}$. The rest of the system parameters take the same values as shown in Table~\ref{tab:supple-parameters across simulations}. In addition, we let the relative mode number $\mu = -2$ in \eqref{eq:supple-lle} and  set the sweeping speed $\mathfrak a = \SI{1.7}{\giga\hertz/\milli\second}$ for the frequency separation $\Omega(t) = \omega_{\text{I, }t=0} + \mathfrak at/2 - \omega_\mu$. The numerical simulation result is summarized as spectrograms displayed in Fig.~\ref{fig:s1-numerical}. These simulation results clearly show strong resemblance to the corresponding experimental data shown in Fig.~4 of the main manuscript.
    \begin{figure}[h]
        \centering
        \includegraphics{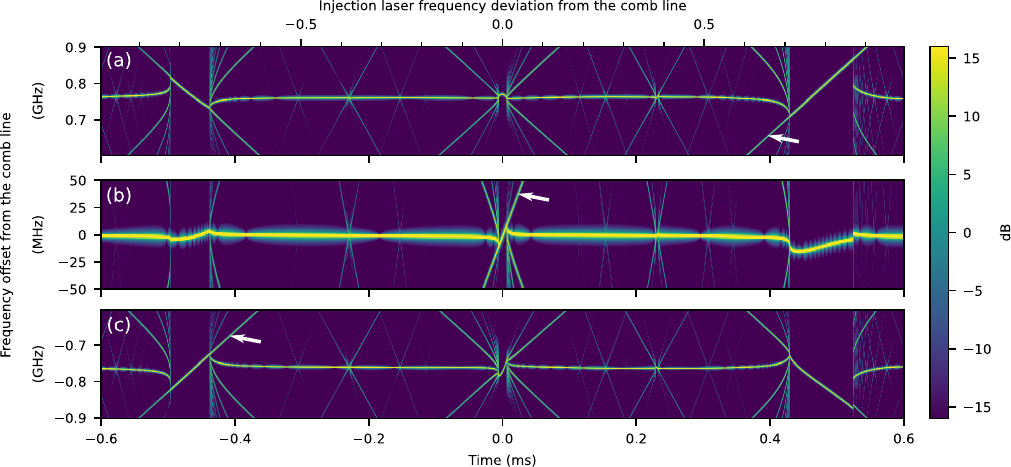}
        \caption{Numerical simulation results corresponding to the spectrograms shown in Fig.~4 in the main text. The white arrows in each panel marks the injection laser frequency component. (a) Frequency variation in the high-frequency breathing sideband. (b) Frequency variation in the $\mu = -2$ comb line. (c) Frequency variation in the low-frequency breathing sideband.}
        \label{fig:s1-numerical}
    \end{figure}

    \section{Derivation of Adler equation from the Poinc\'are normal form}
    \noindent As described in the main text, it is well that soliton breathers in Kerr resonators arise through a Hopf bifurcation. This bifurcation can be expressed in the Poinc\'are normal form as below~\cite[p. 107-108]{kuznetsov2023elements}, which can be derived from the Lugiato-Lefever equation via the technique of central-manifold reduction: 
    \begin{equation}\label{eq:supple-poincare}
        \dv{w}{t} = \lambda w + c_1 |w|^2w,
    \end{equation}
    where the complex variable $w$ represents the state of the system local to the bifurcation point, $c_1$ is some complex constant, and $\lambda$ is the eigenvalue that crosses the imaginary axis. To describe a steady free-running oscillation motion, we assume that $w=r_0\exp(i\omega_{\text{b0}} t)$ in \eqref{eq:supple-poincare}, where $\omega_{\text{b0}}$ corresponds to the free-running (angular) breathing frequency, $\text{d}r_0/\text{d} t = 0$, and we let $c_1 = \alpha + i\beta$, where $\alpha$, $\beta$ are real numbers. This leads to
    \begin{equation}\label{eq:supple-free-running}
        r_0 = \sqrt{-\frac{\lambda'}{\alpha}}, \quad \text{and }\quad \omega_{\text{b0}} = \lambda'' + \beta r_0^2,
    \end{equation}
    where the real numbers $\lambda'$ and $\lambda''$ are respectively the real and imaginary part of the eigenvalue $\lambda$.

    In order to model the synchronization of oscillations, we add a time-dependent perturbation term to the right-hand side of the Poincar\'e normal form~\eqref{eq:supple-poincare} as follows
	\begin{equation}\label{eq:5-1-hopf-inj}
		\dv{w}{t} = \lambda w + c_1|w|^2w + \varepsilon e^{i\varpi_{\text I} t},
	\end{equation}
	where $\varepsilon$ and $\varpi_{\text I}$ are respectively the amplitude and angular frequency of the perturbation term. The perturbation term may be considered as the consequence of the injection laser; however, the reduction of a manifold into the Poincar\'e normal form seems to obscure the mapping of $\varepsilon$ and $\varpi_{\text I}$ to the actual magnitude and frequency of the injection laser radiation. Therefore, care should be taken in the interpretation of the physical meaning associated with $\varepsilon$ and $\varpi_{\text I}$ in \eqref{eq:5-1-hopf-inj}. 
	
	We now assume that the state of the breathing oscillation takes the form of the following ansatz
	\begin{equation}\label{eq:5-1-inj-ansatz}
		w = w(t) = r(t)\exp(i\varpi_{\text I} t + i\phi(t)),
	\end{equation}
	The instantaneous angular breathing frequency $\omega_{\text b}$ can now be calculated from the ansatz~\eqref{eq:5-1-inj-ansatz} as below\enlargethispage{\baselineskip}
	\begin{equation}\label{eq:5-1-omega_b}
		\omega_{\text b} = \varpi_{\text I} + \dv{\phi}{t}.
	\end{equation}
	After substituting the ansatz \eqref{eq:5-1-inj-ansatz} into the modified Poincar\'e normal form \eqref{eq:5-1-hopf-inj} and separating the real and imaginary parts, we get
	\begin{align}
		\dv{r}{t}&=\lambda'r+\alpha r^3 + \varepsilon\cos\phi, \label{eq:5-1-r-eq}\\
		\dv{\phi}{t}&=(\lambda'' + \beta r^2 - \varpi_{\text I}) - \frac{\varepsilon\sin\phi}{r}.\label{eq:5-1-phi-eq}
	\end{align}
	We now assume that $\varepsilon$ is rather small compared to $r_0$ such that we can ignore the term $\varepsilon\cos\phi$ in \eqref{eq:5-1-r-eq}. This small amplitude approximation allows us to set the time derivative of $r$ to zero, and we have $r = r_0$ according to \eqref{eq:supple-free-running}. For the equation \eqref{eq:5-1-phi-eq} of $\phi$, we substitute $r$ with $r_0$ to get (under the small $\varepsilon$ assumption)
	\begin{equation}\label{eq:5-1-pre-adler}
		\dv{\phi}{t} = (\lambda'' + \beta r_0^2 - \varpi_{\text I}) - \frac{\varepsilon\sin\phi}{r_0} = (\omega_{\text{b0}} - \varpi_{\text I}) - \frac{\varepsilon\sin\phi}{r_0},
	\end{equation}
	where in the second equality we use the free-running oscillation formula $\omega_{\text{b0}} = \lambda'' + \beta r_0^2$ from \eqref{eq:supple-free-running}.
	We define $\delta_L = \varpi_{\text I} - \omega_\text{b0}$ and re-write the equation \eqref{eq:5-1-pre-adler} as follows
	\begin{equation}\label{eq:5-1-adler}
		\dv{\phi}{t} =  -\delta_L - \frac{\varepsilon\sin\phi}{r_0}.
	\end{equation}
	The equation in the form shown as \eqref{eq:5-1-adler} is the Adler equation, and it was first obtained by Adler, when investigating the injection locking of electronic oscillators~\cite{adler1946study}.

    We now discuss the locking range and the frequency pulling effect based on the Adler equation~\eqref{eq:5-1-adler}. The locking range is the width of the frequency interval within which synchronization can occur, and we refer to this frequency interval as the synchronization window.	If the synchronization happens, then the instantaneous breathing frequency $\omega_\text b$ is equal to $\varpi_\text I$, or
	\begin{equation}
		\omega_\text b = \varpi_\text I = \varpi_\text I + \dv{\phi}{t},
	\end{equation}
	where \eqref{eq:5-1-omega_b} is used. Consequently, we have $d\phi/dt = 0$, and the equation
	\begin{equation}\label{eq:5-1-adler-state}
		f(\phi) = \delta_L + \frac{\varepsilon\sin\phi}{r_0} = 0
	\end{equation}
	should have solutions of $\phi$. Because the value of $\sin\phi$ ranges from $-1$ to $1$, we can easily see that the synchronization window of $\delta_L$ is as below
	\begin{equation}\label{eq:5-1-lk-delta-range}
		-\frac{\varepsilon}{r_0}\leqslant \delta_L \leqslant \frac{\varepsilon}{r_0}.
	\end{equation}
	Since $\delta_L = \varpi_\text I - \omega_{\text b0}$, we have
	\begin{equation}\label{eq:5-1-lk-omega-range}
		\omega_{\text{b0}} - \frac{\varepsilon}{r_0}\leqslant \varpi_\text I \leqslant \omega_\text{b0} + \frac{\varepsilon}{r_0},
	\end{equation}
	 and the locking range $\Delta\omega_\text b$ can be calculated as follows
	 \begin{equation}\label{eq:5-1-lk-formula}
	 	\Delta\omega_\text b = \Delta\varpi_\text I = \frac{2}{r_0}\varepsilon.
	 \end{equation}
		 The analysis above shows that the synchronization is to be expected over a finite range of frequencies within the synchronization window, and the frequency width of this window, or the locking range, depends linearly on the amplitude $\varepsilon$.
	
	 We shall move on to discuss the phenomenon of the pulling effect. This effect, as described in~\cite{adler1946study}, is most obvious when $\delta_L$ is outside but close to the synchronization window as specified in \eqref{eq:5-1-lk-delta-range}.
	 For the value of $\delta_L$ that is outside the synchronization window \eqref{eq:5-1-lk-delta-range}, we can show by a direct integration using the formula in~\cite[p.\ 99]{dwight1961tables} that the phase $\phi$ satisfies the equation below
	 \begin{equation}\label{eq:5-1-adler-integration}
	 	\delta_L\tan\frac{\phi}{2} + \sqrt{\delta_L^2 - \frac{\varepsilon^2}{r_0^2}}\tan\left(\frac{t}{2}\sqrt{\delta_L^2 - \frac{\varepsilon^2}{r_0^2}} + C \right)  =  -  \frac{\varepsilon}{r_0},\quad\text{when}\quad\delta_L^2\geqslant\frac{\varepsilon^2}{r_0^2}
	 \end{equation}
	 where $C$ is an arbitrary constant of integration. When the phase $\phi = (2N+1)\pi$, where $N\in\mathbb{Z}$, the first term $\tan(\phi/2)$ in \eqref{eq:5-1-adler-integration} becomes infinite. However, the right-hand side of \eqref{eq:5-1-adler-integration} is always finite. This can only happen when the second term on the left-hand side of \eqref{eq:5-1-adler-integration} is also infinite (otherwise the terms do not add up to a finite value), which means we can choose the constant $C$ such that
	 \begin{equation}
	 	t = \frac{(\pm 2N + 1)\pi - 2C}{\sqrt{\delta_L^2 - \varepsilon^2/r_0^2}},\quad\text{when}\quad\phi = (2N+1)\pi\quad\text{and}\quad\delta_L^2\geqslant\frac{\varepsilon^2}{r_0^2}.
	 \end{equation}
	 The time interval during which the phase $\phi$ increases by $2\pi$ can therefore be deduced, and this periodicity leads to an oscillation frequency of~\cite{adler1946study}
	 \begin{equation}\label{eq:5-1-pull1}
	 	\omega_\text b = \varpi_\text I \pm \sqrt{\delta_L^2 - \frac{\varepsilon^2}{r_0^2}}, \quad\text{when}\quad\delta_L^2\geqslant\frac{\varepsilon^2}{r_0^2},
	 \end{equation}
  where $\omega_\text b$ is the pulled breathing frequency. The equation above, when combined with $\delta_L = \varpi_\text I - \omega_{\text{b0}}$ and \eqref{eq:5-1-lk-formula}, gives
  \begin{equation}\label{eq:supple-pulled-angular}
      \omega_\text b = \varpi_\text I \pm \sqrt{(\varpi_\text I - \omega_{\text{b0}})^2 - \left(\frac{\Delta\omega_\text b}{2}\right)^2}
  \end{equation}
  when $\varpi_\text I$ is close to but outside the synchronization window. The non-angular formulation of the above equation gives the pulled breathing frequency formula mentioned in the main text,
  \begin{equation}
      f_\text b = f_\text I \pm \sqrt{(f_\text I - f_{\text{b0}})^2 - \left(\Delta f_\text b/2\right)^2},
  \end{equation}
  where $f_\text b = \omega_\text b/(2\pi)$, $f_\text I = \varpi_\text I/(2\pi)$, $f_\text{b0} = \omega_\text{b0}/(2\pi)$, and $\Delta f_\text b = \Delta\omega_\text b/(2\pi)$.

  \end{widetext}

\end{document}